\documentclass[pre,twocolumn,showpacs,amssymb,floatfix]{revtex4}
\usepackage{graphicx}

\begin{document}

\title{Clustering in a One-dimensional Inelastic Lattice Gas}

\author{Srdjan Ostojic, Debabrata Panja and Bernard Nienhuis}
\affiliation{Institute for Theoretical Physics, Universiteit van
Amsterdam, Valckenierstraat 65, 1018 XE Amsterdam, The Netherlands}

\date{\today}
\begin{abstract} 
We analyze a lattice model closely related to the one-dimensional
inelastic gas with periodic boundary condition. The one-dimensional 
inelastic gas tends to form high density clusters of particles with
almost the same velocity, separated by regions of low density; plotted
as a function of particle indices, the velocities of the gas particles
exhibit sharp gradients, which we call shocks. Shocks and clusters
are seen to form in the lattice model too, although no true positions
of the particles are taken into account. The locations of the shocks
in terms of the particle index show remarkable independence on the
coefficient of restitution and the sequence of collisions used to
update the system, but they do depend on the initial configuration of
the particle velocities. We explain the microscopic origin of the
shocks. We show that dynamics of the velocity profile inside a cluster
satisfies a simple continuum equation, thereby allowing us to study
cluster-cluster interactions at late times.
\end{abstract}


\maketitle

\section{Introduction}

Dynamics of granular fluids has captured a lot of attention from
theoretical physicists for the last few years. In a theorist's model,
the constituent particles of a granular fluid, usually considered to
be hard spheres of finite radii, irretrievably dissipate kinetic
energy via inelastic binary collisions and interparticle frictional
forces. As a result, unlike the microscopic models for the classical
kinetic theory of gases, a granular fluid that is not driven by
external forces ``cools freely''.

Even in the absence of frictional forces, in a stark contrast to hard
sphere fluids with elastic inter-particle binary collisions, such a
simplified model of granular fluids exhibits complex behaviour at
macroscopic scales. In its simplest form, a freely cooling initially
homogeneous and isotropic (both in the particles' position and
velocity space) inelastic gas spontaneously forms nontrivial structures
in the macroscopic velocity  as well as in the macroscopic density
field of the gas. A large number of studies, mostly from the point of
view of inelastic hydrodynamics, have been carried out to understand
the onset for the formation of these structures in two and three
dimensions (Refs. \cite{young,young1,gold,ernst1,brey1,soto,twan} to
cite a few). The (qualitative and quantitative) picture that has
emerged from these studies is that in two and three dimensions, the
system of inelastic hard spheres suffers from inherent long wavelength
linear instabilities. When the system size allows such long
wavelengths to be present, these instabilities start to generate
inhomogeneities in the macroscopic velocity and the density field of
the gas. In the subsequent evolution, these inhomogeneities interact
nonlinearly to give rise to macroscopic structures and the entire
system evolves into a collection of densely populated clusters that
are separated by regions containing particles at very low density
\cite{gold,brey2}. At late times, the clusters collide in a very
complex manner and merge --- a phenomenon that is known as coarsening
in the literature \cite{poschel,luding}.

The late time evolution of a freely cooling inelastic gas is thus
qualitatively completely different from the linear instability
mechanisms at early times. However, although a very large number of
studies have been devoted to kinetic theory of freely cooling inelastic
gases in two and three dimensions, due to the difficulties associated
with the nonlinearities in the behaviour of individual clusters and
cluster-cluster collisions, a proper theoretical understanding of the
long time dynamics of freely cooling inelastic gases has remained
elusive. The existing results have only been numerical
\cite{young1,chen,her,hill,poschel,luding}.

At the other extreme, fully analytical solutions have been found for
completely inelastic (or ``sticky'') granular gas (of point particles)
in one dimension \cite{Frach1}, and it has been shown that the sticky
 gas in one dimension is described by the Burgers equation in
the inviscid limit \cite{Frach1}. In addition, a recent experiment has
also observed clustering in a one-dimensional granular gas
\cite{ko}. Due to dimensional reasons, the dynamics of a granular gas
in one dimension is qualitatively different from those in two or three
dimensions (e.g., vortices cannot form, a strict ordering of particles
from left to right is maintained at all times), but structures are
still seen to form in the velocity and as well as in the density field
of the gas \cite{Kad1,BN}. Ben-Naim et al \cite{BN} have 
studied the formation of these structures, and have conjectured that
at the long times, the behaviour of a one-dimensional inelastic gas
should be the same as that of the sticky gas.

Our purpose in this paper is to unravel some of these long-time
phenomena at a more microscopic level in a simple one-dimensional
lattice model that has been introduced in Ref. \cite{Bald, Bald2}. In this
model, one considers a system of $N$ particles on the integral lattice
positions (denoted by $k$) of a ring of size $N$ with initial
velocities chosen randomly from a uniform distribution in
$[-1,1]$. The ordering of the particles is maintained at all times;
however, since the particles do not move in this lattice model, the
velocities of the particles are not the time derivative of their
positions. Instead, $v_k$, the velocity of the $k$th particle, is
simply a scalar quantity associated with the $k$th particle, and $v_k$
changes only when the $k$th particle participates in a collision with
one of its neighbours according to the following collision rule. At a
collision between $k$th and $(k+1)$th particles at any time, the
post-collisional velocities $v_{k,(k+1)}^{(+)}$ are related to their
pre-collisional velocities $v_{k,(k+1)}^{(-)}$ by
\begin{eqnarray}
v_{k,(k+1)}^{(+)}\,=\,v_{k,(k+1)}^{(-)}\,\pm\,\frac{1+r}{2}\,\left[v_{(k+1)}^{(-)}-v_{k}^{(-)}\right]\,,
\label{e1}
\end{eqnarray}
where $0\leq r\leq1$ is the coefficient of restitution. 
 Time is measured by the average number of
collisions per particle in this lattice model.

\begin{figure}[h]
\begin{center}          
\includegraphics[width=0.83\linewidth]{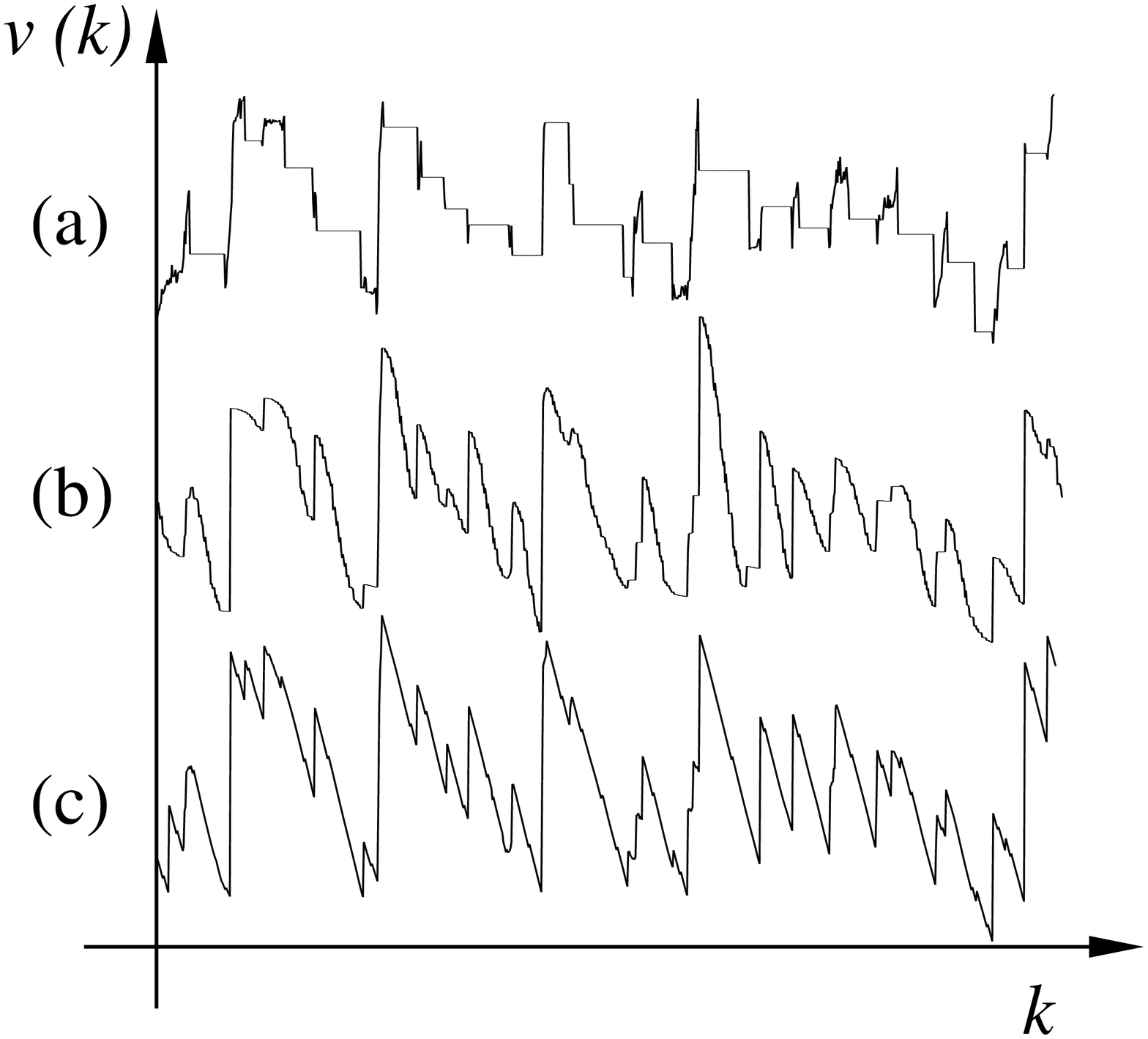}
\caption{Comparison of the shock locations on a ring of size $N$ as a
function of particle positions $k$, $k=1,2,\ldots,N$: (a) a snapshot
of one-dimensional inelastic gas \cite{BN} for $r=0.3$, (b) a
similar snapshot of the random lattice model for $r=0$,(c)a 
snapshot of the systematic lattice model. All systems
had identical initial positions and velocities of the
particles. Visual inspection shows that the locations of the
(relative) large shocks are aligned almost perfectly. The scale in the
$y$-direction is arbitrary.}
\label{twolatt_shocks}
\end{center}
\end{figure}
In this paper, we will consider two variants of this lattice model,
namely the ``random lattice model'' and the ``systematic lattice
model''. Both obey the collision rule (\ref{e1}), but they differ in
the way a colliding pair of particles is chosen to update the
system. For the random lattice model, at any update, the colliding
pair of particles are chosen randomly from all particles that
momentarily satisfy the kinetic constraint $v_{k+1}-v_k<0$. On the other hand, for the
systematic lattice model, the colliding pair of particles are the ones
that has the momentary global minimum value of $v_{k+1}-v_k$. 

The random lattice model {\it without kinetic constraint}, has already
been studied  as part of a larger class of
models, the inelastic Maxwell models \cite{Max}, in which the collision frequency
is chosen to be independent of the incoming velocities of colliding
particles. For that model, it has been analytically shown that
correlations develop with a diffusively growing correlation length,
which consequently affects the temperature decay rate. In addition, the inelastic lattice model in
one-dimension {\it with the kinetic  constraint}, which we study here,
was also analyzed in Refs. \cite{Bald,Bald2} in terms of velocity
distribution and structure factors \cite{remark}.
 It is
however important to realize the difference between these existing
results and the ones reported in this paper: the existing results
mainly concern global quantities, while
in this paper, our main thrust is to study the behaviour of the
microscopic inhomogeneities arising from  the kinetic constraint.



The key feature of both variants of the lattice model with kinetic
constraint studied here is
that an initial configuration of random velocities of the particles
soon develops distinct 
spatial structures,
 eventually leading to {\it large
positive sharp jumps in the particle velocities intersparsed with
relatively smooth variations}. We refer to the large positive sharp
jumps in the velocity field as {\it shocks} and the region between two
consecutive shocks, where the velocity variations are smooth, as {\it
clusters}. These structures have already been observed in \cite{Bald},
but have not been fully analyzed. Formation of shocks and the subsequent dynamics of the
clusters make the lattice model interesting on its own, but its
relevance is realized only when the locations of its shocks in
$v(k)$-profile are compared to those of the one-dimensional inelastic
gas \cite{Frach1,BN} (see Fig. \ref{twolatt_shocks}).
By contrast, there are no shocks in the one dimensional random lattice
 model without kinetic constraints \cite{Max}.

This paper is organized in the following manner: in Sec. \ref{sec2},
we discuss the generic features of the lattice model and explore its
connections with the one-dimensional inelastic gas \cite{Frach1,BN}. In
Sec. \ref{sec3}, we analyze the formation of shocks at early times and
cluster dynamics at late times. We finally end the paper with a short
discussion in Sec. \ref{sec4}.

\section{Generic features of the lattice model\label{sec2}}

\subsection{Phenomenology of shock development and the subsequent
dynamics for the random lattice model\label{sec2a}}

\begin{figure}
\begin{center}              
\includegraphics[width=3.8cm]{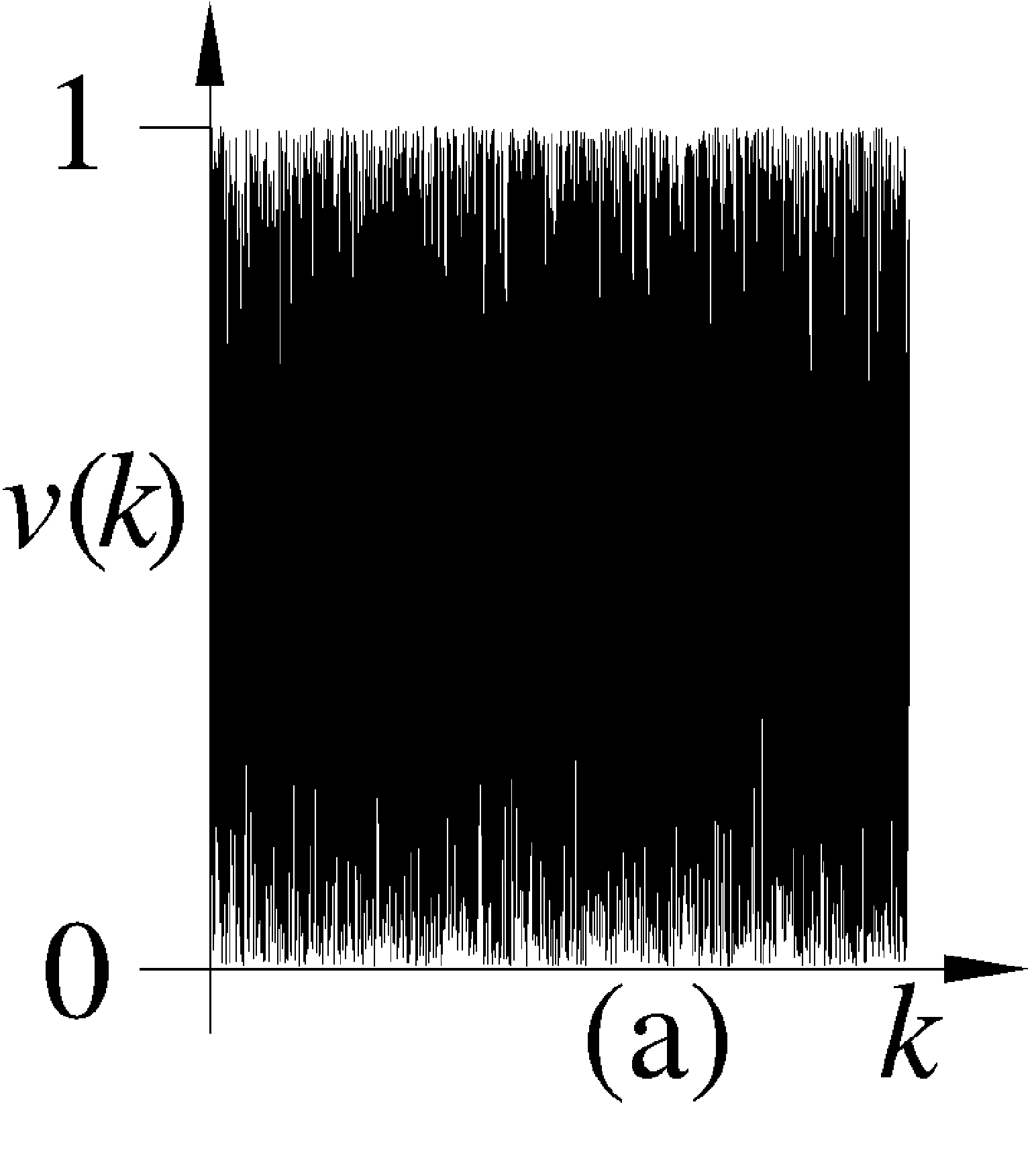}
\hspace*{0.2cm} \includegraphics[width=3.8cm]{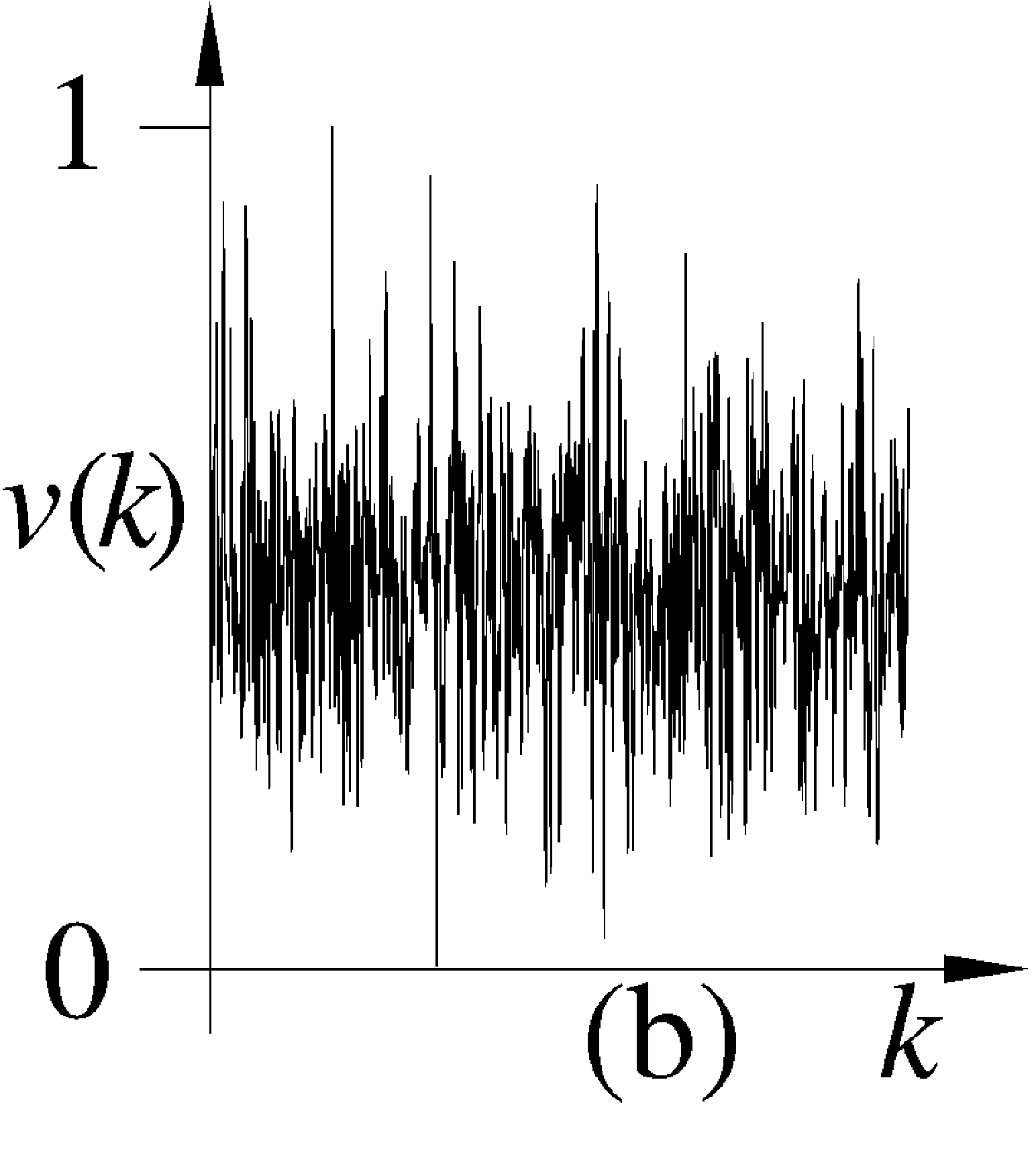}\\
\vspace*{0.2cm} \includegraphics[width=3.8cm]{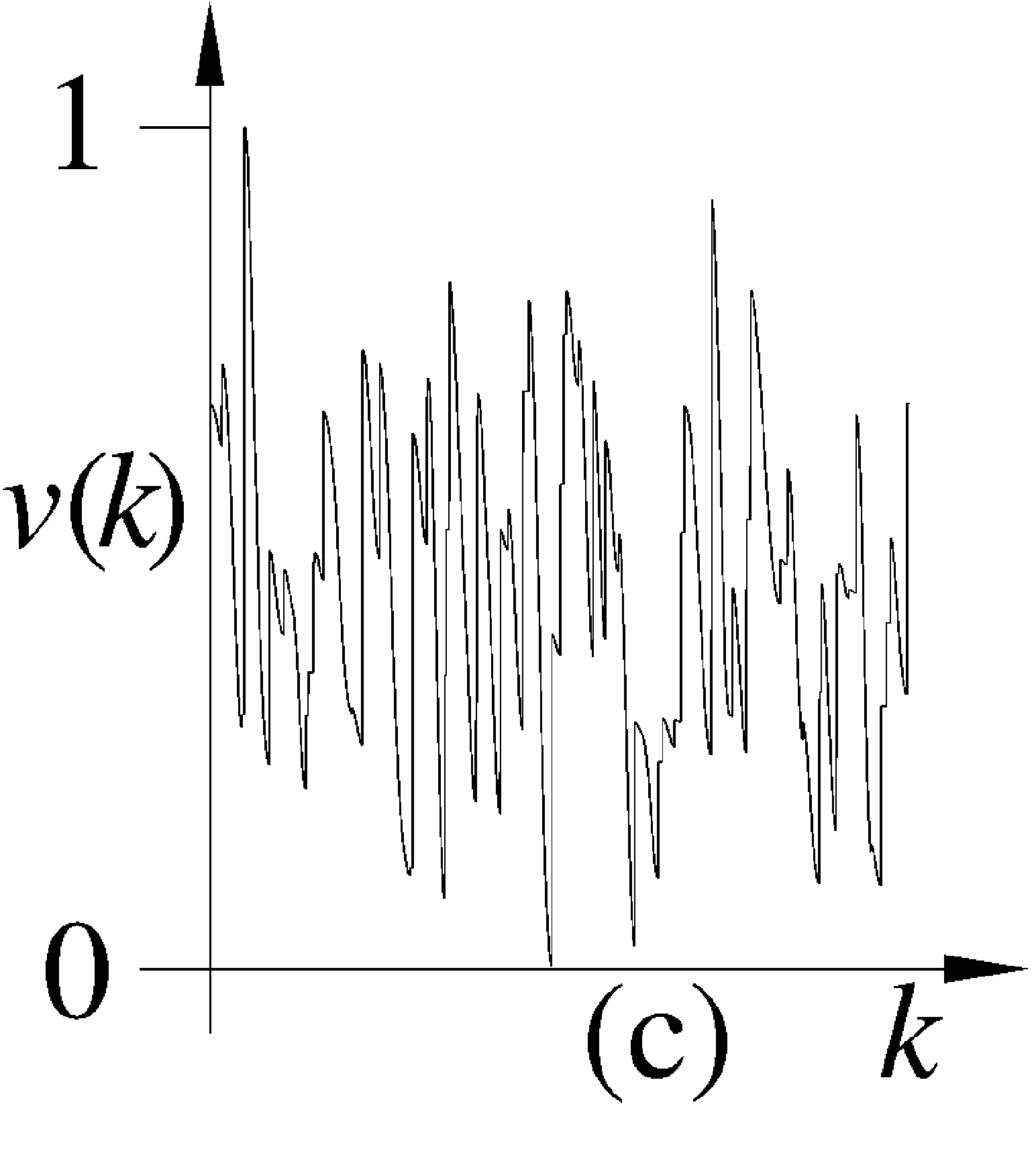}
\hspace*{0.2cm} \includegraphics[width=3.8cm]{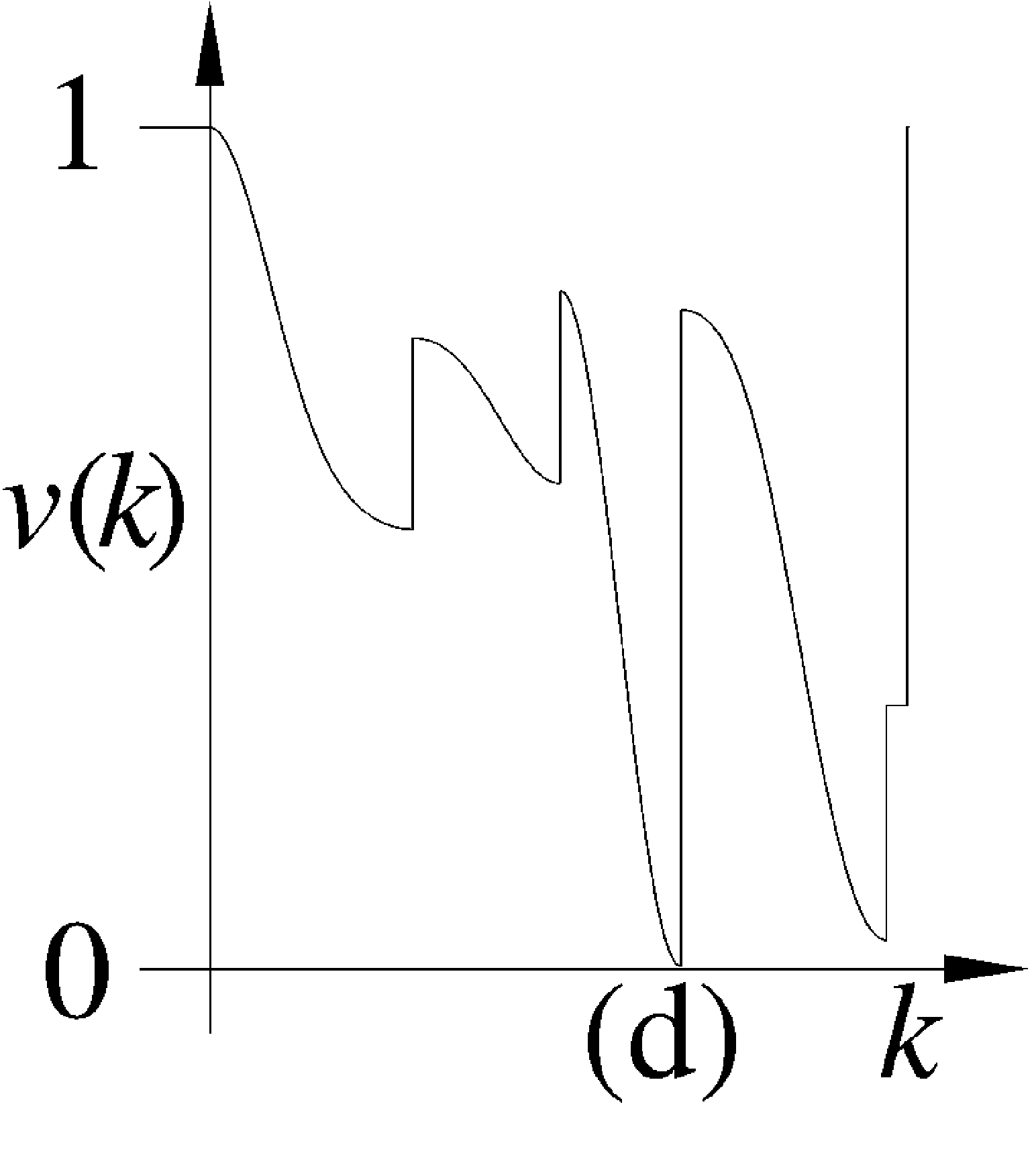}
\end{center}
\caption{An example time sequence of shock developments for a system
of 10000 particles in the random lattice model: (a) Initial profile
(randomly chosen $v_k$ from a uniform distribution in $[-1,1]$), (b)
after 10 collisions per particle, (c) after $10^4$ collisions per
particle, (d) after $10^6$ collisions per particle. The largest
$|v_{k_1}-v_{k_2}|\,\,\forall\,k_1,k_2$ have been scaled to unity in
each graph.}
\label{fig1}
\end{figure}

To start with, in Fig. \ref{fig1}, we show a time sequence of the
velocity profile $v(k)$ for the random lattice model  of $10000$
particles with $r=0.7$. The initial configuration [Fig. \ref{fig1}(a)]
is created by choosing $v_k$ randomly from a uniform distribution in
$[-1,1]$. As can be seen in Fig. \ref{fig1}(b), velocity correlations
set in very rapidly (within $10$ collisions per particle). After
$10^4$ collisions per particle, shocks and clusters can be clearly
identified [Fig. \ref{fig1}(c)]. According to the kinetic constraint, neighbouring particles cannot collide across a shock, and
as a result, each cluster evolves independently of the others, until
two neighbouring clusters collide and coalesce to form a new bigger
cluster [this mechanism is illustrated by the evolution of velocity
profile of Fig. \ref{fig1}(c) to that of \ref{fig1}(d);
Fig. \ref{fig1}(d) corresponds to a state after $10^6$ collisions per
particle].

From the above phenomenological description, one can identify two separate
regimes: the initial homogeneous regime, and the subsequent clustered
regime at late times. To make the evolution of the system from one
regime towards the other quantitatively more precise, we notice that
it corresponds to a symmetry breaking of the probability distribution
$P(\Delta v_k,t)$ of relative velocities $\Delta
v_k=v_{k+1}-v_k$. Initially, for each $k$, the distribution of $v_k$
is uniform in $[-1,1]$ (i.e., a ``box'' distribution). Naturally,
$P(\Delta v_k,t=0)$ is a piecewise linear distribution on $[-2,2]$ (a
``triangular'' distribution), symmetric around $\Delta v_k=0$. Under
the effect of the dynamics, this symmetry is not preserved.  This is
easily understood from the fact that only the neighbouring particle
pairs with $\Delta v_k<0$ can collide, and after a collision between
the $k$th and $(k+1)$th particle, $\Delta v_{k-1}$, $\Delta v_{k}$ and
$\Delta v_{k+1}$ change in a way that, in general, does not preserve
the symmetry of $P(\Delta v_k,t)$.  As this process continues, from
phenomenological considerations, we know that a few shocks remain at
late times. Now, {\it since shocks correspond to large values of
$\Delta v_k$, this means that in the clustered regime, the
probability to have a large positive value of $\Delta v_k$ is
expected to be greater than that of a correspondingly large negative
value}.

To detect the locations of the shocks by following the evolution of
$P(\Delta v_k,t)$ as a function of time, one can identify the location
$k$ of a shock by requiring $\Delta v_k>C\,|\text{min}_j\,\Delta v_j|
$, where  $C>0$ is a constant. The choice of the numerical value of
$C$ is arbitrary, and this arbitrariness can be used to tune the
minimum value that $\Delta v_k$ must have in order to qualify for a
shock. 
 Furthermore, if one
defines the instant $t_p$ when the system passes from one regime to
the other by the minimal time where at least one shock becomes
visible, then the above requirement can also be used to characterize
$t_p$. Of course, the precise value of $t_p$ depends on the chosen
value of $C$.

In the clustered regime, due to inelastic collisions, the amplitude of
the velocity profile within each cluster decreases, and the velocities
of the particles approach that of the center of mass of the cluster
itself. This process continues for a while until two neighbouring
clusters, with center of mass velocities $V_j$ and $V_{j+1}$ collide
and coalesce and the shock separating them disappears [at time $t$, if
we number the clusters $j=1,2,\ldots,j_1(t)$ from the left to the
right, then such a cluster-cluster collision takes place only if
$V_j>V_{j+1}$]. Precisely this mechanism is responsible for making the
velocity profile of Fig. \ref{fig1}(c) evolve to that of
Fig. \ref{fig1}(d).

At very late times, due to periodic boundary condition, only a single
cluster survives. Its amplitude also decreases in time, and the
velocities of all particles asymptotically converge to the center of
mass velocity of the whole system.

\subsection{Dependence on the coefficient of restitution for the
random lattice model \label{sec2b}}

One of the first questions that one can ask for the lattice model is
what is the effect of inelasticity on the clustering properties of the
system. Clearly, for $r=1$, the collisions are elastic, and upon
collision, the velocities of two neighbouring particles are simply
interchanged. Such a dynamics cannot lead to clustering or any
structure formation.

\begin{figure}[h]
\begin{center}              
\includegraphics[width=0.45\linewidth]{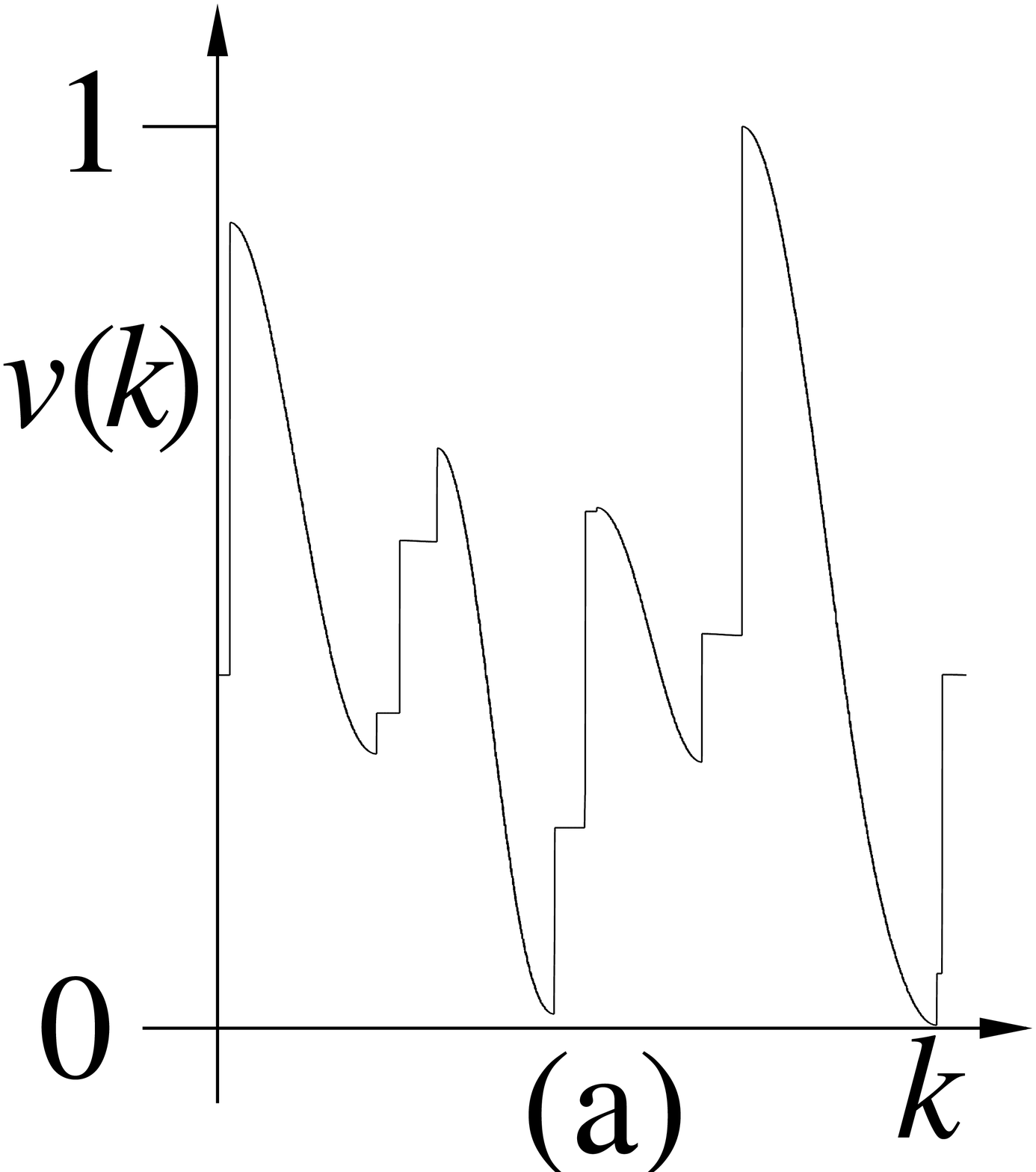}
\hspace{0.2cm} \includegraphics[width=0.45\linewidth]{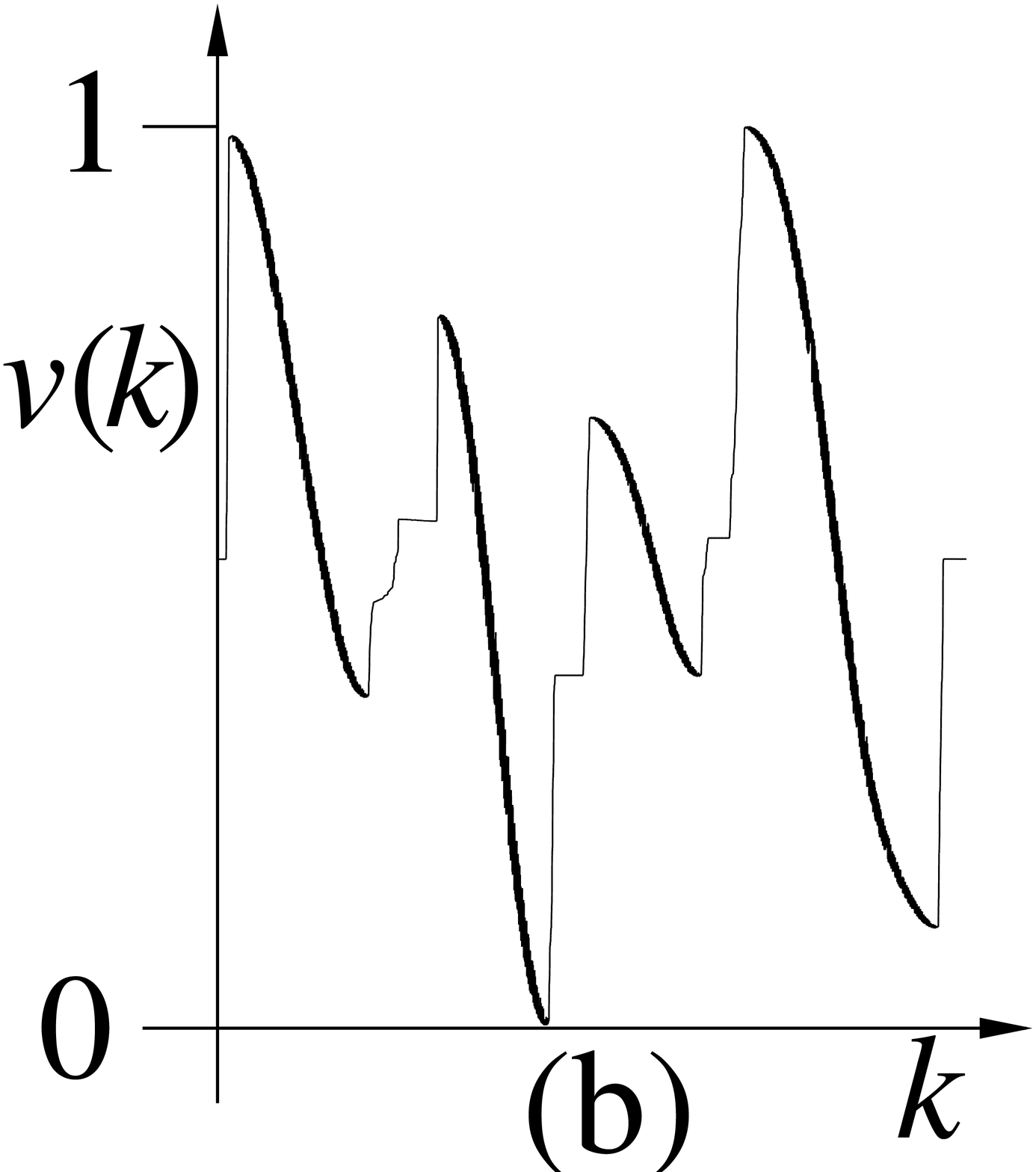}
\end{center}
\caption{Velocity profiles for the random lattice model of $5000$
particles: (a) for $r=0.7$ after $10^4$ collisions per particle (b)
for $r=0.95$ and $10^3$ collisions per particle.  The initial velocity
configurations of the particles were the same for both (a) and
(b). Once again, the largest $|v_{k_1}-v_{k_2}|\,\,\forall\,k_1,k_2$
has been scaled to unity for both profiles.}
\label{rex}
\end{figure}

On the other hand, our simulations show that the lattice model
exhibits clustering for any value of $r$ smaller then $1$, and it is
only $t_p$ defined in Sec. \ref{sec2a} that becomes an increasing
function of $r$. The most remarkable observation, however, is that for
a given initial velocity configuration of the particles, not only  the
locations of the shocks persisting at late times are the same for all
$r$, but also the macroscopic velocity profiles are nearly independent
of $r$.  If we consider the macroscopic velocity profile of the
lattice model with a coefficient of restitution $r_1$ at a time $t_1$
well into the clustered regime, then for any $r_2>r_1$, (originating
from the same initial velocity configuration of the particles) there
exists a time $t_2<t_1$, for which the two macroscopic velocity
profiles are almost the same (see Fig. \ref{rex}). Effectively, this
implies that for all $r$, the dynamical behaviour of the system is
identical to that for $r=0$ at late times. We also observe that at
intermediate times, the maximum number of shocks (and correspondingly,
clusters) in the system are observed for $r=0$ (see Sec. \ref{sec3a}
in this regard).

\subsection{Dependence on the order of collisions and on the initial
velocity configuration of the particles\label{sec2c}}

The next important question one can ask is if not $r$, exactly what
influences the positions of the shocks? A priori, one expects that the
locations of the shocks and the long time dynamics of the system
depend on the initial velocity configuration of the particles,  as
well as on the sequence in which the collisions are performed.

Surprisingly however, the locations of the shocks depend {\it only\/}
on the initial configuration of the particles' velocities, e.g, two
different random collision sequences result in the same locations of
the shocks. On the other hand, the variant of the lattice model
(random or systematic) determines the macroscopic shape of the
velocity profile inside the clusters. To demonstrate this phenomenon,
starting from the same initial velocities of the particles, we
simulated both the random and the systematic lattice models and
compared the locations of the shocks (see Fig. \ref{twolatt_shocks}).

\subsection{Relation between the lattice model and one-dimensional
inelastic gas\label{sec2d}}

The empirical observations in the preceding sections suggest that so
long as one is only interested in the velocity profile $v(k)$, a close
relation exists between the one-dimensional inelastic gas and the
lattice model from the following consideration: the dynamics of a
one-dimensional inelastic gas for a given initial configuration is
completely deterministic --- after any  collision, the next colliding
pair is automatically determined by the instantaneous  minimal value
of $(x_{k+1}-x_k)/(v_{k}-v_{k+1})$, for $v_{k}-v_{k+1}>0$. In this
sense, one can view the one-dimensional inelastic gas as a variant of
the lattice model too, where the colliding particles are chosen in a
very complicated manner.  Section \ref{sec2c} then indicates that for
a given initial configuration of the particles' velocities as a
function of the particle index, the locations of the shocks in the
one-dimensional inelastic gas and those in the lattice model are the
same. That this is indeed the case is numerically verified in
Fig. \ref{twolatt_shocks}.

In our simulations of the one-dimensional inelastic gas, we followed
the procedure outlined in Ref. \cite{BN}, namely that to avoid the
inelastic collapse, the collisions were taken to be elastic when the
relative velocities between the neighbouring particles are smaller
than a given cutoff. Our results confirm the existence of ``Burgers''
shocks in the $v(x)$ profile \cite{BN}. More importantly, we note that
for the one-dimensional inelastic gas, {\it there is an inverse
correspondence of shocks and clusters between this $v(x)$ and $v(k)$
profile --- each shock in the $v(x)$ profile [i.e., a large particle
density $n(x)$] corresponds to a cluster in the $v(k)$ profile and the
shocks in the $v(k)$ profile correspond to regions with finite
gradients in $v(x)$}.

\section{Analysis\label{sec3}}

\subsection{Development of shocks\label{sec3a}}

The lesson that we learnt from Sec. \ref{sec2} is that the locations
of the shocks in this lattice model are essentially determined from
the initial configuration of the particles' velocities. This
immediately gives rise to the following question: how can one predict
the locations of the shocks seen at early times from the initial
velocity configuration of the particles?

\begin{figure}[h]
\begin{center}          
\includegraphics[width=0.6\linewidth]{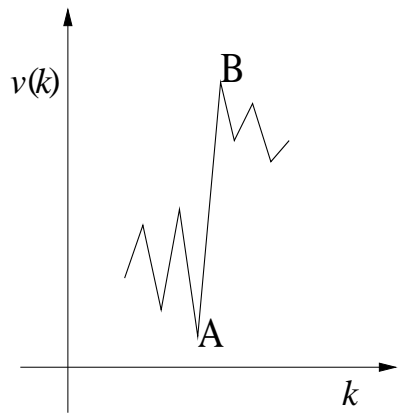}
\caption{A schematic diagram of the initial velocity profile for a few
particles to explain the development of shocks.}
\label{ab}
\end{center}
\end{figure}
As it turns out, most of the shocks observed in the clustered regime
already exist at $t=0$. These are the locations marked by large
positive jumps in the $v(k)$ profile between two neighbouring lattice
sites at $t=0$, surrounded by relatively small velocity variations. To
illustrate this point, we present a schematic diagram for such a
velocity profile for a few particle in Fig. \ref{ab}. In the
subsequent dynamics at short times, the particles on the right of B
and on the left of A collide with each other and ``thermalize'' (i.e.,
the velocity variations reduce due to inelasticity), but this
thermalization may not necessarily convert a relatively large
$v_B-v_A>0$ to $v_B-v_A<0$ quickly enough to make the A and B
collide. In other words, such a large positive jump remains preserved
and may eventually give rise to a shock.

\begin{figure}[h]
\begin{center}              
\includegraphics[width=0.9\linewidth]{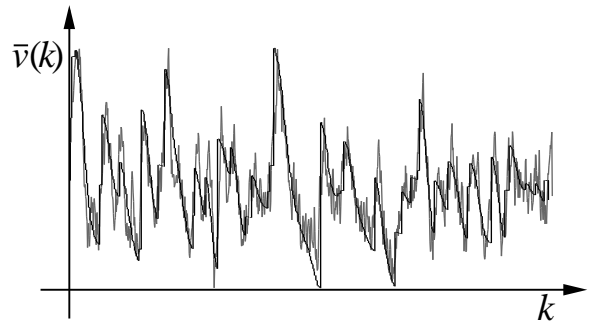}
\caption{Correspondence between the coarse grained velocity profiles
$\bar{v}(k)$ ($n=7$) for $r=0$ for the random lattice model of $1000$
particles. Gray curve corresponds to $t=0$ and the black curve
represents early shocks. The scale in the $y$-direction is arbitrary.}
\label{cg}
\end{center}
\end{figure}

To check if such a scenario is correct or not, one needs to identify
the locations of such large positive jumps in the initial $v(k)$
profile, and contrast them with the observed locations of the shocks
at early times. A convenient way to single out such jumps [say, in
Fig. \ref{fig1}(a)] is to define a $(2n+1)$-coarse grained velocity
profile $\bar{v}(k)=\displaystyle{\sum_{i=k-n}^{k+n}}v_i/(2n+1)$ of
the particles, which, roughly speaking, can single out one such large
jump over the surrounding small velocity variations in a window of
$(2n+1)$ lattice sites in the $\bar{v}(k)$ profile. Having fine-tuned
$n$, a surprisingly close match between the observed shock locations
at early times and the initial configuration of the particles'
velocities can be found (see Fig. \ref{cg}).

Notice, however, in Fig. \ref{cg} that not all the large jumps in the
initial configuration of the particles' velocities have turned out to
be shocks at a later time. Indeed, in general, whether such an initial
large positive jump at a particular location develops into a shock or
not really depends on the magnitude of the jump itself in relation to
particle velocities in its immediate vicinity, and as well as on the
coefficient of restitution $r$. In this regard, the dependence on the
coefficient of restitution is not difficult to understand, as $r$
quantifies the ``transport of velocity'' between two colliding
particles. Intuitively speaking, an inelastic collision between two
neighbouring particles can be viewed as a combination of dissipation
(thermalization) and transport of velocities. For $r=1$ only transport of
velocities can take place. At the other limit, for $r=0$, there is no
transport of velocities but only dissipation. As a result, for $r=0$,
small velocity variations around any large jump thermalize
immediately, leading to an early appearance of the clustered regime,
and a large number of initial jumps end up becoming shocks. With
increasing $r$, some of the initial large positive jumps are
eliminated by transport of velocities, and clustering regime, with a
smaller number of selected shocks (and clusters), appears later.

The appearance of the shocks at the locations where relatively large
positive jumps exist is therefore caused by the thermalization of the
smaller velocity variations around these jumps --- due to
thermalization, the smaller velocity variations become even smaller,
but the large positive jumps remain preserved under the dynamics. In
relation to their surroundings, the magnitude of these large jumps
thus starts to grow in time. Such a scenario applies independently of
the details of the model, and gives a qualitative explanation for the
observation that starting from a given initial configuration, shocks
appear at the same locations in the $v(k)$ profile for both variants
of the lattice model and as well as for the one-dimensional inelastic
gas \cite{Frach1,BN}. Moreover, it also clearly demonstrates that the
shock formation ``instability'' in the $v(k)$ profile in all these
models is a ``relative instability'' and not an ``absolute'' one. Such
behaviour has also been found in the (long-wavelength) linear
instability of the macroscopic flow field for two- and
three-dimensional inelastic gases (see Refs. \cite{gold,ernst1,twan}
for example).

\subsection{Cluster dynamics\label{sec3b}}

\subsubsection{Dynamics within a single cluster\label{sec3b1}}

Once the shocks develop, our simulations show that for a given variant
of the lattice model, the velocity profile within each cluster has the
same characteristic macroscopic shape. For the deterministic lattice
model, this shape is simply linear with a negative slope. Although we
cannot provide an analytical derivation of this shape, if we interpret
the dynamics along the lines that at each updating step of a cluster,
two neighbouring particles collide at the location where the
$v_{k+1}-v_k$ is minimum (which is a mechanism that tries to align all
$v_{k+1}-v_k$ values to its maximum value within each cluster), then
the linear profile appears to be intuitively reasonable.

For the random lattice model, however, the slope of the macroscopic
velocity profile is smoothly varying, as can be seen in, e.g.,
Fig. \ref{fig1}(d) [occasionally, flat profiles  can be observed too,
but for the time being, we leave them for Sec. \ref{sec3b2}]. It turns
out that the functional form of the smooth macroscopic velocity
profile can be obtained through a mean-field approach by averaging
over an ensemble of random collision sequence realizations within a
cluster, as we describe below.

The idea behind this mean-field approach is the following: we consider
a cluster of $M$ particles, and denote the velocity of the $k$th
particle at time $t$, averaged over random collision sequences, by
$\langle v_k(t)\rangle$.  From the very definition of a cluster in the
introduction, we assume that $\langle v_{k+1}(t)\rangle-\langle
v_k(t)\rangle<0$ at all times. Thereafter, as we observe from direct
simulation measurements that the probability of collisions between any
two neighbouring particle pairs are equally likely, in this mean field
approach, $\langle v_k(t)\rangle$ is easily seen to satisfy the
equation
\begin{eqnarray}
\langle v_k(t+1)\rangle\,=\,(1-2p)\,\langle
v_k(t)\rangle\nonumber\\&&\hspace{-3cm}+\,p\,[\langle
v_k(t)\rangle\,-\,\varepsilon\{\langle v_k(t)\rangle-\langle
v_{k-1}(t)\rangle\}]\nonumber\\&&\hspace{-3cm}+\,p\,[\langle
v_k(t)\rangle\,+\,\varepsilon\{\langle v_{k+1}(t)\rangle-\langle
v_k(t)\rangle\}].
\label{mf}
\end{eqnarray}
Here, $p$ is an effective probability of a collision between any two
neighbouring particles in this mean-field theory and
$\varepsilon=(1+r)/2$. The three terms on the r.h.s. of
Eq. (\ref{esol}) respectively originate from the events when the $k$th
particle is not involved in a collision, when there is a collision
between the $k$th and the $(k-1)$th particle and when there is a
collision between the $k$th and the $(k+1)$th particle. As
Eq. (\ref{mf}) very simply reduces to
\begin{eqnarray}
\langle v_k(t+1)\rangle\,=\,(1-2p\varepsilon)\,\langle
v_k(t)\rangle\,+\,p\varepsilon \,\langle
v_{k-1}(t)\rangle\nonumber\\&&\hspace{-3cm}+\,p\varepsilon\,\langle
v_{k+1}(t)\rangle\,,
\label{emean}
\end{eqnarray} 
the interesting point to note is that Eq. (\ref{emean}) is the
discrete (both in space and time) form of the diffusion equation
\begin{eqnarray}
\frac{\partial \langle v(k,t)\rangle }{\partial
t}\,=\,D(r)\,\frac{\partial^2 \langle v(k,t)\rangle}{\partial k^2}\,,
\label{e2}
\end{eqnarray}
where mean-field theory predicts $D(r)=p\varepsilon$. {\it One has to
keep in mind however that Eq. (\ref{e2}) holds only for monotonically
decreasing $\langle v(k,t)\rangle$ as a function of $k$}. A similar
equation has also been found  in the one-dimensional lattice model
without the kinetic constraint \cite{Max}.

To solve Eq. (\ref{e2}), the macroscopic velocity profile of a cluster
at time $t_0$ can be generally expanded in a Fourier series. Since
particle velocities are not transported across the boundaries of a
cluster, $\displaystyle{\frac{\partial\langle v_k(t)\rangle}{\partial
k}}$ must vanish at the boundaries of a cluster, the Fourier series
contains only terms $\propto\cos[\pi jk]$. The amplitude of the $j$th
such term decreases as $e^{-\pi^2j^2D(r)(t-t_0)}$, so that for large
values of $(t-t_0)$, only the slowest decay mode $j=1$ survives.
Hence at long times we expect  $\langle v(k,t)\rangle$ to be given by
\begin{eqnarray}
\langle v(k,t)\rangle\,=\,A(t)\cos\left[\frac{\pi
k}{M}\right]\nonumber\\&&\hspace{-3.05cm}=\,A(t_0)\,e^{-\pi^2D(r)(t-t_0)/M^2}\cos\left[\frac{\pi
k}{M}\right]\,.
\label{esol}
\end{eqnarray}
\begin{figure}[h]
\begin{center}              
\includegraphics[width=4cm,height=3.8cm]{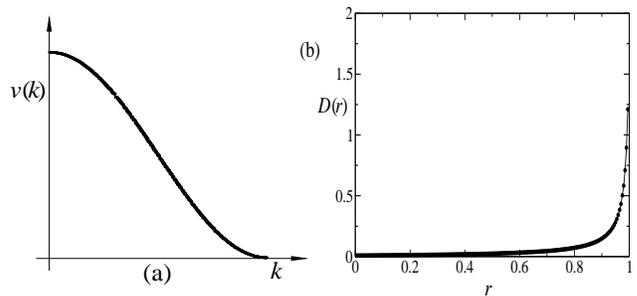}\hspace*{-0.3cm}
\includegraphics[width=4.5cm,height=4cm]{Figs/rdep}
\end{center}
\caption{(a) Comparison of the half-cosine solution (\ref{emean})
[solid line] and simulation data[open circles], the two are completely
indistinguishable from each other, (b) Numerically obtained dependence
of $D(r)$ on $r$ for the random lattice model. The simulations were
carried out for $M=1000$. The scale along the $y$ axis for Fig. (a) is
arbitrary. $D(r)$ exhibits a singularity at $r=1$, which is in
clear disagreement with the mean-field theory prediction.}
\label{halfcos}
\end{figure}
 
The mean-field result (\ref{esol}) allows us to compare the
half-cosine form of $\langle v(k,t)\rangle$ with the macroscopic
velocity profile of the particles within a cluster, as observed in the
simulation for the random lattice model. This comparison is shown in
Fig. \ref{halfcos}(a) ---  the fact that we cannot distinguish the
simulation data from the half-cosine shape of $\langle v(k,t)\rangle$
as predicted by Eq. (\ref{esol}) is an indication of how well the
mean-field approach works to describe the average shape of an isolated
cluster. In addition, by following the dynamics from the simulation,
we can also obtain an empirical functional form of $D(r)$ [shown in
Fig. \ref{halfcos}(b)].

There are two features of the $D(r)$ vs. $r$ curve that require
further elaboration. The first one of them is that $D(r)$ is an
increasing function of $r$, indicating that the amplitude of the
cluster $A(t)$ decreases faster as $r$ increases. This observation
seems to contradict the fact that the dissipation decreases with
increasing $r$. The point to notice however is that the mean-field
approach really describes the time evolution within a cluster in which
$\displaystyle{\frac{\partial\langle v_k(t)\rangle}{\partial
k}}<0$. From that point of view, the amplitude of the cluster
decreases not only from the dissipation, but also from transport of
velocities. This is easily seen from the fact that after a collision,
the ordering of the magnitudes of the velocities of the colliding
particles are simply exchanged, i.e., the particle on the right has a
higher velocity, which results in a local positive slope in the $v(k)$
profile. For increasing $r$, transport of velocities become more
effective, and thus the negative slope of
$\displaystyle{\frac{\partial\langle v_k(t)\rangle}{\partial k}}$
within a cluster decays to zero faster. Only if $r=1$, the velocities
of the particles within an isolated cluster order themselves in an
increasing order of magnitude. However, $D(r)$ has a singularity at
$r=1$ and the $r=1$ case cannot be treated within the scope of our
mean-field theory. Secondly, from the way we presented our mean-field
approach, it may seem that the
$r$-dependence of $D(r)$ can be obtained through the dependence of $p$
and $\varepsilon$ on $r$. This is actually not true. The mean-field
approach (\ref{mf}) neglects fluctuations in the particles' velocities
around $\langle v(k,t)\rangle$, but in an actual simulation, these
fluctuations are very important as they decide the sign of $\Delta
v_k(t)=v_{k+1}(t)-v_k(t)$ for the particles within the cluster and
thereby control which collisions are possible and which are not.

\subsubsection{Interacting clusters for the random lattice
model\label{sec3b2}}

With the background of Sec. \ref{sec3b1}, it is now clear that as time
progresses, the amplitude of each cluster present in the random
lattice model at late times effectively decays exponentially as
$\sim\exp[-\pi^2D(r)(t-t_0)/M^2]$. Such a decay brings the velocities
of the particles within a cluster closer and closer to the
centre-of-mass velocity of the cluster itself. Let us assume that at
time $t_0\gg1$, there are $j_1$ number of clusters present in the
whole system of the random lattice model. If we number these clusters
by $j$ such that $j=1,2,\ldots,j_1$ and denote the number of particles
within the clusters, the amplitudes of the clusters and their
velocities of their centre of masses respectively by $M_j$, $A_j$ and
$V_j$, then with increasing time, we observe the following dynamics:
(i) when $V_{j-1}<V_j<V_{j+1}$, the shocks on the left and on the
right of the $j$th cluster cannot respectively be smaller than
$V_j-V_{j-1}>0$ and $V_{j+1}-V_j>0$. In that case, the amplitude $A_j$
simply decreases to $0$, forming a ``flat'' velocity profile (see
e.g., Fig. \ref{rex}). (ii) On the other hand, when $V_j>V_{j+1}$, the
magnitude of the shock between the $j$th and the $(j+1)$th clusters
decreases to zero, and the two clusters coalesce together to form a
new bigger cluster in a finite time. For a given configuration of
clusters at time $t$, the mean-field theory of Sec. \ref{sec3b1}
provides us with a way to measure the time $t'=t+\Delta t_{min}$ of
the first coalescence of two clusters in the system. In fact, $\Delta
t_{min}=\text{min}_j\,\,\Delta t_j$, where for each pair of
neighbouring clusters $(j,j+1)$ with $V_j-V_{j+1}>0$, $\Delta t_j$ is
obtained by solving the equation
\begin{eqnarray}
A_{j+1}(t_0)\,\exp[-\pi^2D(r)\Delta
t_j/M_{j+1}^2]\nonumber\\&&\hspace{-5cm}+\,A_{j}(t_0)\,\exp[-\pi^2D(r)\,\Delta
t_j/M_{j}^2]\,=\,V_j-V_{j+1}.
\label{e3}
\end{eqnarray}

Due to the fact that Eq. (\ref{e3}) is transcendental in nature, a
closed form analytical solution for $\Delta t_j$ is impossible to
obtain, let alone the value of $\Delta t_{min}$. Nevertheless,
Eq. (\ref{e3}) provides us with a glimpse of how complicated it is to
theoretically study the cluster-cluster collisions and coalescence in
the random lattice model, and in general, in inelastic gases. If two
clusters $l$ and $l+1$ are the first ones to coalesce (at time
$t_0+\Delta t_1$) in the random lattice model, then a new cluster with
$M_l+M_{l+1}$ particles and center of mass velocity $(M_l V_l+M_{l+1}
V_{l+1})/(M_l+M_{l+1})$ is formed. At time $t_0+\Delta t_1$, the shape
of the new cluster is different from a half-cosine, but as the
mean-field theory of Sec. \ref{sec3b1} suggests, very soon the shape
of the new cluster converges to a half-cosine, unless the newly formed
cluster collides with another one in the meantime.

\subsubsection{A ``hydrodynamic'' description of the random lattice
model\label{sec3b3}}

So far, we have analyzed the system-wide properties of shocks and
clusters. We have also seen that there exists an effective dynamics in
terms of diffusion equations (\ref{e2}) and (\ref{esol}) within a
single cluster. Based on this collected wisdom on the random lattice model
so far, one can naturally ask if it is possible to express the
system-wide properties of shocks and clusters in terms of an effective
(``hydrodynamic'') equation.

It turns out that indeed such an equation can be constructed for the
random lattice model:
\begin{eqnarray}
\frac{\partial v}{\partial t}\,=\,D(r)\,[\,\Theta(-\nabla^+ v)\,\nabla^+v\,-\Theta(-{\nabla}^{-} v)
\,\nabla^-v]\,,
\label{Sidedif}
\end{eqnarray}
where $\Theta$ denotes a unit step function,
$(\nabla^+v)_k=v_{k+1}-v_{k}$ and
$(\nabla^-v)_k=v_{k}-v_{k-1}$ are the discrete gradients
operating to the right and left respectively and
$\displaystyle{\frac{\partial}{\partial t}}$ the discrete time
gradient. It is important to stress that in a numerical implementation
of Eq. (\ref{Sidedif}),  all pairs of particles with negative relative
velocities collide {\it simultaneously\/} in a unit timestep. The main
difference between the numerical implementation of Eq. (\ref{Sidedif})
and the random lattice model lies in the order of the collisions, but
in the light of the previous results, we expect to find the shocks at
the same locations. Moreover, inside a given cluster, (\ref{Sidedif})
reduces to (\ref{e2}), so that the cluster velocity profiles should be
the same as in the random lattice model.
\begin{figure}[h]
\begin{center}              
\includegraphics[width=0.6\linewidth]{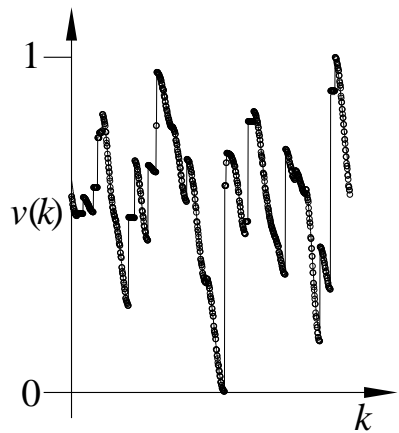}
\end{center}
\caption{Comparison of the macroscopic velocity profile (solid curve)
predicted by Eq. (\ref{Sidedif}) with the corresponding actual
simulation data (open circles) of the random lattice model for
$r=0$. The simulations were carried out for $1000$ particles and the
largest $|v_{k_1}-v_{k_2}|\,\,\forall\,k_1,k_2$ have been scaled to
unity both for the actual simulation data and the solid curve.}
\label{lastfig}
\end{figure}

Indeed, we find that the macroscopic velocity profile obtained
numerically from Eq. (\ref{Sidedif}) compares very well with the
actual computer simulation results of the random lattice model (see
Fig. \ref{lastfig}), as expected.

\section{Discussion\label{sec4}}

In this paper, we have extensively studied structure formation and
their dynamics in a one-dimensional inelastic lattice model for
granular gases. These structures appear in the form of clusters
separated by shocks. The locations of the shocks in this lattice model
at early times are decided from the initial configurations of the
particles' velocities in a very robust manner. In order to predict the
locations of the shocks, we have demonstrated a procedure to process
the initial velocity configurations data of the particles, and this
procedure works equally well also for the one-dimensional inelastic gas
\cite{Frach1,BN}. The coefficient of restitution does play a role to
decide which of the large jumps in the particles' velocities at early
times yield shocks, but at late times, the macroscopic velocity
profile for a given model with a given initial configuration of the
particles' velocities is independent of the coefficient of
restitution.

Thus, we observe that in terms of detailed structure
formation, systems with any $r<1$ ``flow'' towards the sticky limit
$r=0$. Such ``universality'' has been found in Ref. \cite{BN} in terms
of global quantities, while our results suggest that a broader
``universality''
 holds for microscopic quantities, such as
the locations of shocks (when the same initial velocity configuration
is considered).

In addition, for the random lattice model, we have also studied the
dynamics of an isolated cluster and cluster-cluster collisions in
detail  and obtained an effective ``hydrodynamic'' equation. We hope
that the analyses presented here can be successfully used to study
coarsening problems in realistic granular gases (i.e., in two or three
dimensions).

S.O. and D.P. are financially supported by the Dutch research
organization FOM (Fundamenteel Onderzoek der Materie).


\begin{thebibliography}{03}

\bibitem{gold} I. Goldhirsch, M.-L. Tan and G. Zanetti,
Phys. Rev. Lett. {\bf 70}, 1619 (1993); J. Sci. Comp. {\bf 8}, 1
(1993).

\bibitem{young} S. McNamara and W. R. Young, Phys. Rev. E {\bf 50},
R28 (1994).

\bibitem{young1} S. McNamara and W. R. Young, Phys. Rev. E {\bf 53},
5089 (1996).

\bibitem{ernst1} J. A. G. Orza, R. Brito, T. P. C. Van Noije and M. H.
Ernst, Int. J. Mod. Phys. C {\bf 8}, 953 (1997).

\bibitem{brey1} J. Javier Brey, F. Moreno and M. J. Ruiz-Montero,
Phys. Fluids {\bf 10}, 2965 (1998); {\bf 10}, 2976 (1998).

\bibitem{soto} R. Soto, M. Mareschal and M. Malek-Mansour,
Phys. Rev. E {\bf 62}, 3836 (2000).

\bibitem{twan} T. P. C. van Noije and M. H. Ernst, in {\it Granular
Gases}, eds. T. Poschel and S. Luding (Springer, NY, 2001).

\bibitem{brey2} J. Javier Brey, M. J. Ruiz-Montero and D. Cubero,
Phys. Rev. E {\bf 60}, 3150 (1999).

\bibitem{poschel} T. P\"{o}schel and S. Luding eds., {\it Granular
Gases}, Lecture Notes in Physics 564 (Springer, Berlin, 2001).

\bibitem{luding} S. Miller and S. Luding, {\it Cluster Growth in Two-
and Three-dimensional Granular Gases}, cond-mat/0304637.

\bibitem{chen} S. Chen, Y. Deng, X. Nie and Y. Tu, Phys. Lett. A {\bf
269}, 218 (2000).

\bibitem{her} S. Luding and H. J. Herrmann, Chaos {\bf 9}, 673 (1999).

\bibitem{hill} S. Hill and G. F. Mazenko, Phys. Rev E {\bf 67}, 061302
  (2003)

\bibitem{Frach1} L. Frachebourg,  Phys. Rev. Lett.  {\bf 82}, 1502
(1999); L. Frachebourg, Ph. A. Martin, J. Piasecki, Physica A {\bf
279}, 69 (2000); L. Frachebourg, Ph. A. Martin, J. Fluid
Mech. {\bf417}, 323 (2001)

\bibitem{ko} R. Mikkelsen, M. Versluis, E. Koene, G.-W. Bruggert,
D. van der Meer, L. van der Weele, and D. Lohse, Phys.  Fluids {\bf
14}, S14 (2002), \verb+ http://ojps.aip.org/phf/gallery/pdf/2002/S14_1.pdf+.

\bibitem{Kad1} L. P. Kadanoff, Rev. Mod. Phys. {\bf 71}, 435 (1999)

\bibitem{BN} E. Ben-Naim,S. Y. Chen, G. D. Doolen and S. Redner,
Phys. Rev. Lett.  {\bf 83}, 4069 (1999).

\bibitem{Bald} A. Baldassarri, U. Marini Bettolo Marconi, A. Puglisi,
Europhys. Lett. {\bf 58}, 14 (2002).

\bibitem{Bald2} A. Baldassarri, A. Puglisi, U. Marini Bettolo Marconi,
{\it ``Kinetic Models of Inelastic Gases''}, Lecture Notes in Physics
624, 95 (2003) (cond-mat/0302418).

\bibitem{Max} E. Ben-Naim, P. L. Krapivsky, {\it ``The Inelastic Maxwell
Model''}, Lecture Notes in Physics 624, 65 (2003) (cond-mat/0301238).

\bibitem{remark}Note that  this model does not belong to the Maxwell family since the collision frequency does depend on the
incoming velocities.
\end{thebibliography}
\end{document}